\documentclass[aps,prl,reprint,superscriptaddress,longbibliography,nofootinbib]{revtex4-2}
\usepackage[T1]{fontenc}
\usepackage{lmodern}
\usepackage{amsmath,amssymb,mathtools,bm}
\usepackage{graphicx}
\usepackage{microtype}
\usepackage[colorlinks=true,linkcolor=blue,citecolor=blue,urlcolor=blue]{hyperref}
\newcommand{\Var}{\operatorname{Var}}
\newcommand{\Prob}{\mathbb P}
\newcommand{\E}{\mathbb E}
\newcommand{\meanOmega}{\langle\Omega\rangle}
\newcommand{\dd}{\mathrm d}
\newcommand{\sgn}{\operatorname{sgn}}
\begin{document}

\title{Bounds for Apparent Second-Law Violations in Quantum Trajectories}
\author{Domingos S. P. Salazar}
\affiliation{Unidade de Educa\c{c}\~ao a Dist\^ancia e Tecnologia, Universidade Federal Rural de Pernambuco, 52171-900 Recife, Pernambuco, Brazil}
\date{August 9, 2026}

\begin{abstract}
 Negative stochastic entropy production is commonly called an apparent violation of the second law.  In general quantum-trajectory dynamics, however, the physical entropy production $\sigma$ need not obey a forward detailed fluctuation theorem.  A general arbitrary-coupling formulation identifies a dynamical-asymmetry term $\sigma^\ast$ that completes it into $\Omega=\sigma+\sigma^\ast$, whose mean is $\meanOmega=\Sigma+\Sigma^\ast$.  We prove that the likelihood-ratio sign is optimal among reversal-odd trajectory observables and use this fact to transfer an established sharp fluctuation-theorem floor to the tie-corrected physical sign statistic $\Pi_\sigma=\Pr(\sigma<0)+\Pr(\sigma=0)/2$.  When $\Pr(\sigma=0)=0$, the result reads $\Pr(\sigma<0)\ge[1-\meanOmega/g(\meanOmega)]/2$, where $g$ is the inverse of $a\mapsto a\tanh(a/2)$.  The physical integral fluctuation theorem simultaneously suppresses large negative events, producing a quantitative ``frequent but mild'' law, while the sign imbalance lower-bounds the hidden mean $\Sigma^\ast$.  We formulate the measured-record protocol explicitly and illustrate and numerically audit the tie-corrected theorem in random finite-coupling collision models and a coherently driven qubit interacting with thermal ancillas.
\end{abstract}
\maketitle

\paragraph{Introduction.---}
Quantum trajectories resolve an open quantum evolution into individual measurement records $\gamma$ \cite{Dalibard1992,DumZollerRitsch1992,MolmerCastinDalibard1993,GisinPercival1992,Carmichael1993,PlenioKnight1998,Brun2002,WisemanMilburn2010,JacobsSteck2006,BreuerPetruccione2002}.  A general finite-coupling construction realizes such records through sequential interactions with fresh environments and two-point measurements \cite{VuHonmaSaito2026}.  The trajectory and repeated-interaction ideas themselves have a broader lineage in quantum stochastic thermodynamics and collision models \cite{LandiPaternostro2021,Ciccarello2022,Manzano2018,Roadmap2026}, while jump, weak-measurement, and continuous-monitoring descriptions assign heat, work, and entropy production to individual records \cite{Esposito2010,Breuer2003,Horowitz2012,HorowitzParrondo2013,HorowitzSagawa2014,HekkingPekola2013,AlonsoLutzRomito2016,Elouard2017,ElouardMohammady2018,ManzanoZambrini2022}.  Following the arbitrary-coupling construction of Ref.~\cite{VuHonmaSaito2026}, we write the completed trajectory log-likelihood as $\Omega(\gamma)=\ln[P(\gamma)/P(\gamma')]$, where $\gamma'$ is the time-reversed record.  The physical entropy production is $\sigma(\gamma)=\ln[P(\gamma)/\widetilde P(\gamma')]$, comparing the forward record with its counterpart in the physically backward experiment, while $\sigma^\ast(\gamma)=\ln[P(\gamma)/\widetilde P(\gamma)]$ quantifies the same-record mismatch between forward and backward dynamics.  Writing $\Sigma=\langle\sigma\rangle$ and $\Sigma^\ast=\langle\sigma^\ast\rangle$, the associated microreversibility identity gives
\begin{equation}
 \Omega(\gamma)=\sigma(\gamma)+\sigma^\ast(\gamma),\qquad
 \meanOmega=\Sigma+\Sigma^\ast.
 \label{eq:completion}
\end{equation}
Thus $\Omega$ is a completed entropy production whose mean contains both physical dissipation and the forward--backward dynamical-asymmetry cost.

The map $\gamma\mapsto\gamma'$ is an involution, $(\gamma')'=\gamma$.  Defining $P'(\gamma)=P(\gamma')$, Eq.~\eqref{eq:completion} implies $\Omega=\ln(P/P')$ and $D_{\rm KL}(P\Vert P')=\meanOmega$.  Let $g$ be the inverse on $[0,\infty)$ of $a\mapsto a\tanh(a/2)$.  Identifying record reversal with the involution used in the iTUR \cite{SalazariTUR2022}, Ref.~\cite{VuHonmaSaito2026} obtained its quantum-trajectory thermodynamic form
\begin{equation}
 \frac{\Var J}{\langle J\rangle^2}\ge
 \operatorname{csch}^2\!\left[\frac{g(\meanOmega)}{2}\right],
\end{equation}
 for every reversal-odd current $J(\gamma')=-J(\gamma)$ with $\langle J\rangle\ne0$.  This precision bound is therefore controlled by the mean completed entropy $\meanOmega$, not by $\Sigma$ alone, and belongs to a broader family of current-based entropy-production inference bounds \cite{Manikandan2020,VuVoHasegawa2020}.

\emph{First main result.---} We ask: \emph{how does $\meanOmega$ control apparent violations of the physical second law?}  Define $\Pi_\sigma=\Prob(\sigma<0)+\Prob(\sigma=0)/2$ and analogously $\Pi_\Omega$; the half weight is unbiased tie breaking at zero.  We write $L(\meanOmega):=[1-\meanOmega/g(\meanOmega)]/2$, with $L(0):=1/2$ by continuity.  The event $\sigma<0$ is negative physical entropy production; $\Omega<0$ instead means that a forward record is less probable than its reversed record under the same forward law.  We prove below the thermodynamic transfer $\Pi_\sigma\ge\Pi_\Omega$; the exact detailed fluctuation theorem frequency bound \cite{SalazarApparent2021} supplies the established floor $\Pi_\Omega\ge L(\meanOmega)$.  Hence $\Pi_\sigma\ge\Pi_\Omega\ge L(\meanOmega)$ is the lower branch of our main result.  Since $\Prob(\sigma\le0)\ge\Pi_\sigma$, its simplest zero-inclusive form is
\begin{equation}
 \Prob(\sigma\le0)\ge
 \frac12\left[1-\frac{\meanOmega}{g(\meanOmega)}\right].
 \label{eq:main}
\end{equation}
The bound is sharp.  Detailed equality conditions and an explicit cyclic-qutrit collision model that attains the floor are given in the Supplemental Material~\cite{SupplementalMaterial}.  Close to reversibility, $L(\meanOmega)=1/2-\sqrt{\meanOmega/8}+O(\meanOmega^{3/2})$, so negative physical-entropy records cannot become rare faster than the square root of the completed dissipation.

Negative trajectory entropy has been the operational content of fluctuation theorems since their original formulation \cite{EvansCohenMorriss1993,GallavottiCohen1995,EvansSearles2002,Jarzynski1997,Crooks1999,VanZonCohen2003,VanZonCohen2004,Jarzynski2006,EspositoHarbolaMukamel2009,Campisi2011,Seifert2005,Seifert2012}, and has been measured in colloids, driven two-level systems, biomolecules, and single-electron devices \cite{Wang2002,Carberry2004,Schuler2005,Tietz2006,Speck2007,Koski2013,Ciliberto2017}.  Quantum experiments now reconstruct work, heat, entropy production, and statistical arrows of time along individual monitored trajectories \cite{Murch2013,Pekola2013,Batalhao2014,Batalhao2015,Brunelli2018,Harrington2019,Naghiloo2020,Rossi2020,Jayaseelan2021,Aguilar2022}.  Related work bounds the probability or duration of second-law-violating fluctuations and optimizes probabilistic work extraction \cite{GarciaGarcia2014,Cavina2016,Maillet2019,MillerPerarnauLlobet2023}; these tail, duration, and work-success questions differ from the sign-frequency floor at fixed $\meanOmega$.  Recent studies have also emphasized that second-law-defying records need not be rare in nonstationary processes and examined them in hybrid quantum--classical systems \cite{Liao2025,Paraguassu2025}.  A distinct open-system literature concerns temporarily negative entropy-production rates associated with memory and system--environment correlations \cite{Bhattacharya2017,Marcantoni2017,Popovic2018,StrasbergEsposito2019}; those rate effects differ from the fixed-horizon event probability controlled here.  Equation~\eqref{eq:main} gives the compact zero-inclusive form even when the physically meaningful $\sigma$ does not itself satisfy a forward detailed fluctuation theorem because $\Sigma^\ast\ne0$.  If $\Prob(\sigma=0)=0$, it reduces to the negative-event probability $\Prob(\sigma<0)$.

\paragraph{Formalism.---}
All likelihood ratios are understood on the common support of the paired measures; records with $P(\gamma)=P(\gamma')=0$ are omitted.  A singular reversal pair gives $\meanOmega=+\infty$ and the trivial limiting floor $L(+\infty)=0$.  Consider a system interacting sequentially with fresh environments.  Let $K_{\mu\nu}=\langle\mu|U|\nu\rangle$ be the conditional system operator associated with an ancilla entering in the energy state $|\nu\rangle$ and leaving in $|\mu\rangle$, and let $M_{\mu\nu}=\sqrt{q_\nu}K_{\mu\nu}$ be the measured Kraus operator \cite{BreuerPetruccione2002}.  For
$\gamma=\{n,(\nu_1,\mu_1),\ldots,(\nu_N,\mu_N),m\}$,
\begin{equation}
 P(\gamma)=r_n\prod_{i=1}^{N}q_{\nu_i}
 \left|\langle m|K_{\mu_N\nu_N}\cdots K_{\mu_1\nu_1}|n\rangle\right|^2.
\end{equation}
Its reversal is $\gamma'=\{m,(\mu_N,\nu_N),\ldots,(\mu_1,\nu_1),n\}$.  The forward and physically backward laws obey \cite{VuHonmaSaito2026}
\begin{equation}
 \widetilde P(\gamma')\widetilde P(\gamma)=P(\gamma)P(\gamma'),
 \qquad p_\Omega(\omega)=e^\omega p_\Omega(-\omega).
 \label{eq:microdft}
\end{equation}
The first identity proves Eq.~\eqref{eq:completion}; the second identity is the Detailed Fluctuation Theorem (DFT) for $\Omega$ and follows by pairing each record with its reverse.  Pathwise derivations of the completion and DFT, together with the connection to the quantum-trajectory iTUR, are provided in the Supplemental Material~\cite{SupplementalMaterial}.

The DFT also gives a short derivation of the completed-event frequency floor.  With $A=|\Omega|$,
\begin{equation}
 \meanOmega=\E\!\left[A\tanh\frac A2\right],\qquad
 \Pi_\Omega=\frac12\left[1-\E\!\left(\tanh\frac A2\right)\right].
\end{equation}
Set $\psi(y)=\tanh[g(y)/2]$.  Writing $a=g(y)$ gives
$\psi''(y)=-\{2\cosh^2(a/2)[h'(a)]^3\}^{-1}<0$, so Jensen's inequality yields
$\E[\tanh(A/2)]\le\psi(\E[h(A)])=\meanOmega/g(\meanOmega)$ and therefore $\Pi_\Omega\ge L(\meanOmega)$ \cite{SalazarApparent2021}.  Equality requires $A=g(\meanOmega)$ almost surely; the binary DFT law with probabilities proportional to $e^{\pm g(\meanOmega)/2}$ realizes this condition.

It remains to compare the sign of $\Omega$ with that of the physical entropy.  More generally, let $X(\gamma')=-X(\gamma)$ and set $\sgn(0)=0$.  Since
\begin{align}
 \langle\sgn X\rangle
 &=\frac12\sum_\gamma [P(\gamma)-P'(\gamma)]\sgn X(\gamma)\nonumber\\
 &\le\frac12\sum_\gamma|P(\gamma)-P'(\gamma)|
 =\langle\sgn\Omega\rangle,
\end{align}
we have $\Pi_X=[1-\langle\sgn X\rangle]/2\ge\Pi_\Omega$.  Microreversibility makes $\sigma$ reversal odd, proving the stated ordering; Eq.~\eqref{eq:main} follows because $\Prob(\sigma\le0)\ge\Pi_\sigma$.  Orbitwise, $\Omega<0$ always selects the less probable member of each pair $\{\gamma,\gamma'\}$; the negative sign of any other odd variable must select either that member, the more probable member, or split the pair when the variable vanishes.

The excess physical-violation probability is exactly
\begin{equation}
 \begin{aligned}
 \Pi_\sigma-\Pi_\Omega
 &=\frac14\sum_\gamma|P(\gamma)-P'(\gamma)|\\[-0.2ex]
 &\quad\times[1-\sgn\sigma(\gamma)\sgn\Omega(\gamma)]\ge0.
 \end{aligned}
 \label{eq:gap}
\end{equation}
It is the distinguishability-weighted penalty for sign disagreement between physical and completed entropy production.  Equality holds when their signs agree on every asymmetric reversal orbit.  Applying the same theorem to $-\sigma$ gives
\begin{equation}
 1-L(\meanOmega)\ge
 \Prob(\sigma<0)+\tfrac12\Prob(\sigma=0)
 \ge L(\meanOmega).
 \label{eq:twosided}
\end{equation}
This is our main result in its complete two-sided form.  Its lower branch implies Eq.~\eqref{eq:main} because $\Prob(\sigma\le0)=\Pi_\sigma+\Prob(\sigma=0)/2\ge\Pi_\sigma$; when $\Prob(\sigma=0)=0$, the two statements coincide as $\Prob(\sigma<0)\ge L(\meanOmega)$.  The DFT further gives $\Pi_\Omega\le1/2$.  The physical $\sigma$ generally obeys an integral fluctuation theorem, and a $1/2$ ceiling does not follow from oddness, the product identity, and that IFT alone.  The valid upper branch $1-L$ is operationally useful near reversibility and in the sign witness below; physical sharpness of this branch under additional protocol constraints remains open.  The full sign-optimality proof, tie treatment, upper-bound analysis, and equality characterization appear in the Supplemental Material~\cite{SupplementalMaterial}.

\begin{figure*}[!t]
\begin{minipage}[t]{0.492\textwidth}
\centering\textbf{(a)}\\[-0.4ex]
\includegraphics[width=\linewidth]{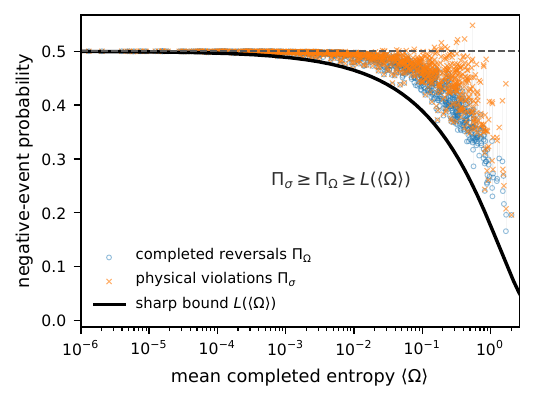}
\end{minipage}\hfill
\begin{minipage}[t]{0.492\textwidth}
\centering\textbf{(b)}\\[-0.4ex]
\includegraphics[width=\linewidth]{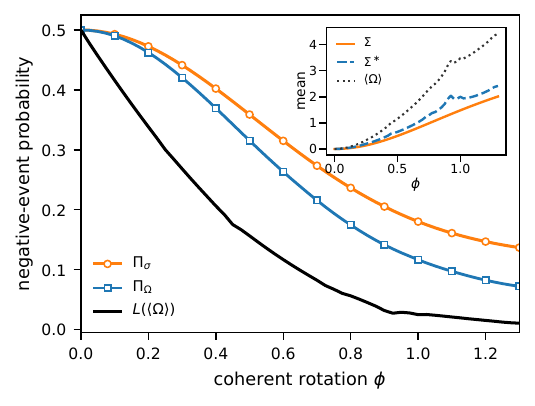}
\end{minipage}
\caption{Sign-frequency bounds.  Here $\Pi_X:=\Prob(X<0)+\Prob(X=0)/2$ for any stochastic quantity $X$; thus $\Pi_\sigma$ is the tie-corrected frequency of apparent second-law violations and equals the literal negative-event frequency when $\Prob(\sigma=0)=0$.  (a) For 1000 random finite-coupling models governed by Eq.~\eqref{eq:randomH}, $\Pi_\sigma$ (orange crosses) lies above $\Pi_\Omega$ (blue circles) and the sharp curve $L(\meanOmega)$; light segments connect the two statistics for the same model.  (b) Exact four-collision statistics for the driven qubit in Eqs.~\eqref{eq:coherentU} and \eqref{eq:coherentpath}.  The inset reports $\Sigma$ and $\Sigma^\ast$ across the drive sweep.}
\label{fig:frequency}
\end{figure*}

\paragraph{Applications.---}
\emph{Measured repeated-interaction protocol.---}
Repeated interactions provide a discrete realization of a monitored quantum trajectory in which the measured record is explicit at every step \cite{Ciccarello2022}.  Each environment $E_i$ is prepared in
$\tau_E=\sum_\nu q_\nu|\nu\rangle\langle\nu|$, with
$q_\nu=e^{-\beta\epsilon_\nu}/Z$, and interacts with the system through the same joint unitary $U$.  The conditional system operator is
$K_{\mu\nu}=\langle\mu|U|\nu\rangle$, while the collision channel is
$\mathcal E(\rho)=\sum_{\mu\nu}q_\nu K_{\mu\nu}\rho K_{\mu\nu}^\dagger$.
We prepare its stationary state
$\rho_{\rm ss}=\sum_n r_n|n\rangle\langle n|$ and projectively measure the system in this eigenbasis before the first and after the final collision.  Every ancilla is measured in the energy basis immediately before and after its interaction.  A record therefore contains the system endpoints $n,m$ and the energy transitions $\nu_i\to\mu_i$ of all ancillas.

For compactness define the two conditional products
\begin{equation}
 \begin{aligned}
 \mathcal K_\gamma
 &=K_{\mu_N\nu_N}\cdots K_{\mu_1\nu_1},\\
 \mathcal K_{\gamma'}
 &=K_{\nu_1\mu_1}\cdots K_{\nu_N\mu_N}.
 \end{aligned}
\end{equation}
The same measured record then determines
\begin{equation}
 \begin{aligned}
 \sigma(\gamma)
 &=\ln\frac{r_n}{r_m}
 +\sum_{i=1}^{N}\ln\frac{q_{\nu_i}}{q_{\mu_i}},\\
 \sigma^\ast(\gamma)
 &=\ln\frac{|\langle m|\mathcal K_\gamma|n\rangle|^2}
 {|\langle n|\mathcal K_{\gamma'}|m\rangle|^2},\\
 \Omega(\gamma)&=\sigma(\gamma)+\sigma^\ast(\gamma).
 \end{aligned}
\end{equation}
The first term in $\sigma$ is the stochastic change of system entropy and the thermal ratios give the entropy flow,
$\ln(q_{\nu_i}/q_{\mu_i})=\beta(\epsilon_{\mu_i}-\epsilon_{\nu_i})$.
The second line is the forward--reverse conditional-amplitude mismatch.  It need not be measured through a physically backward experiment: the same forward histogram supplies
$\Omega(\gamma)=\ln[P(\gamma)/P(\gamma')]$, after which
$\sigma^\ast=\Omega-\sigma$.  Thus the experimental inputs to the main tie-corrected ordering are the endpoint populations, the ancilla energy changes, and paired frequencies of a record and its reversal.  Complete measurement sequences and record-probability formulas are collected in the Supplemental Material~\cite{SupplementalMaterial}.

\emph{Random finite-coupling ensemble.---}
The first application deliberately leaves the perturbative regime.  A qubit collides once with a thermal ancilla of dimension $d_E=2,\ldots,5$ under
\begin{equation}
 \begin{aligned}
 H_{\rm tot}&=H_S\otimes I+I\otimes H_E
 +\lambda V_S\otimes V_E,\\
 H_S&=\tfrac12(\sigma_z+w_x\sigma_x),\\
 H_E&=\sum_\nu\epsilon_\nu|\nu\rangle\langle\nu|,
 \qquad U=e^{-i\tau H_{\rm tot}}.
 \end{aligned}
 \label{eq:randomH}
\end{equation}
The interactions $V_S,V_E$ are independent random Hermitian operators normalized to unit operator norm.  The transverse term in $H_S$ and the generic interaction ensure that neither energy conservation nor commutativity is imposed.  For each realization we solve
$\mathcal E(\rho_{\rm ss})=\rho_{\rm ss}$ and enumerate the forward and reversed-forward probabilities
\begin{equation}
 \begin{aligned}
 P_\gamma
 &=r_nq_\nu|\langle m,\mu|U|n,\nu\rangle|^2,\\
 P'_\gamma
 &=r_mq_\mu|\langle n,\nu|U|m,\mu\rangle|^2,
 \end{aligned}
\end{equation}
so that
\begin{equation}
 \sigma_\gamma=\ln\frac{r_nq_\nu}{r_mq_\mu},
 \qquad
 \Omega_\gamma=\ln\frac{P_\gamma}{P'_\gamma}.
\end{equation}
All $4d_E^2$ records are retained.  In system-gap units we draw
$w_x\in[0.02,0.35]$ and ancilla energies in $[0,1.5]$; the inverse temperature, coupling, and duration are sampled log-uniformly over
$\beta\in[0.25,3]$, $\lambda\in[0.15,7]$, and
$\tau\in[0.25,6]$.  Figure~\ref{fig:frequency}(a) contains 1000 independent realizations and therefore probes the bound across weak and strong collisions, several ancilla dimensions, and noncommuting dynamics without changing the measurement prescription.  The ensemble construction and numerical consistency checks are documented in the Supplemental Material~\cite{SupplementalMaterial}.

Figure~\ref{fig:frequency}(a) shows three relevant features.  All realizations obey $\Pi_\sigma\ge\Pi_\Omega\ge L(\meanOmega)$; a strict orange--blue gap in 741 cases makes sign disagreement common in this ensemble.  Scatter at fixed $\meanOmega$ shows that the black curve is a universal floor, not a prediction of the frequency.  Finally, 17 realizations have $\Pi_\sigma>1/2$ (maximum $0.5481$), giving explicit collision examples in which the DFT ceiling for $\Pi_\Omega$ does not transfer to the physical statistic.

\begin{figure*}[!t]
\begin{minipage}[t]{0.492\textwidth}
\centering\textbf{(a)}\\[-0.4ex]
\includegraphics[width=\linewidth]{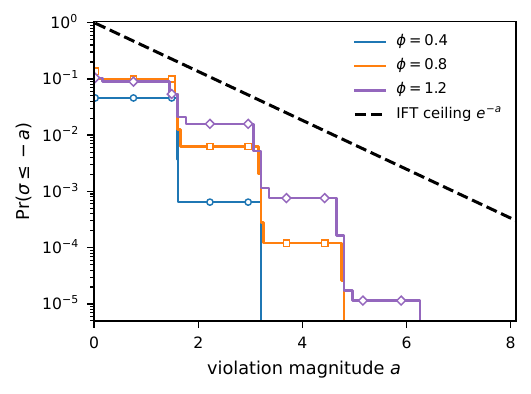}
\end{minipage}\hfill
\begin{minipage}[t]{0.492\textwidth}
\centering\textbf{(b)}\\[-0.4ex]
\includegraphics[width=\linewidth]{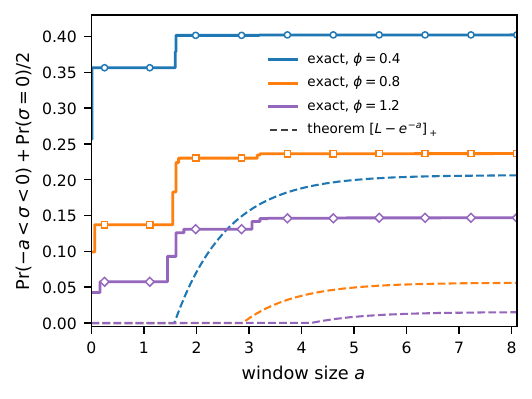}
\end{minipage}
\caption{Tie-corrected frequency and severity.  (a) Exact negative tails and the standard IFT/Markov baseline $e^{-a}$.  (b) The exact quantity $\Prob(-a<\sigma<0)+\Prob(\sigma=0)/2$ (solid) and the combined frequency--severity bound \eqref{eq:mild} (dashed).  All distributions are exact enumerations of the driven four-collision records in Eq.~\eqref{eq:coherentpath}.}
\label{fig:mild}
\end{figure*}

\emph{Coherently driven qubit collisions.---}
The second application resolves how a controlled drive generates the separation between physical and completed entropy.  A system qubit receives an $x$ pulse and then exchanges one excitation with a fresh thermal qubit.  Setting $\hbar=1$ and
$H_S=H_E=\omega|1\rangle\langle1|$, the two stages are generated in the interaction picture by
\begin{equation}
 H_{\rm d}=\frac{\Omega_R}{2}\sigma_x,
 \qquad
 H_{\rm ex}=g_c(\sigma_+\otimes\sigma_-+\sigma_-\otimes\sigma_+),
\end{equation}
with one-step unitary
\begin{equation}
 \begin{aligned}
 U_{\rm ex}(\theta)
 &=e^{-i\theta(\sigma_+\otimes\sigma_-+\sigma_-\otimes\sigma_+)},\\
 U_{\phi,\theta}
 &=U_{\rm ex}(\theta)[R_x(\phi)\otimes I],
 \qquad R_x(\phi)=e^{-i\phi\sigma_x/2}.
 \end{aligned}
 \label{eq:coherentU}
\end{equation}
where $\phi=\Omega_Rt_d$ and $\theta=g_ct_c$.  Writing
$c_\theta=\cos\theta$ and $s_\theta=\sin\theta$, the four energy-resolved conditional operators are
\begin{equation}
 \begin{aligned}
 K_{00}&=(|0\rangle\langle0|+c_\theta|1\rangle\langle1|)R_x,\\
 K_{10}&=-is_\theta|0\rangle\langle1|R_x,\\
 K_{01}&=-is_\theta|1\rangle\langle0|R_x,\\
 K_{11}&=(c_\theta|0\rangle\langle0|+|1\rangle\langle1|)R_x.
 \end{aligned}
 \label{eq:coherentKraus}
\end{equation}
The ancilla weights are
$p_0=(1+e^{-\beta\omega})^{-1}$ and
$p_1=e^{-\beta\omega}p_0$.  We obtain the stationary preparation from
$\rho_{\rm ss}=\sum_{\mu\nu}p_\nu K_{\mu\nu}\rho_{\rm ss}K_{\mu\nu}^\dagger$ and measure its eigenstates at the two endpoints.  Let
$\mathcal K_\gamma^{(\phi)}$ and
$\mathcal K_{\gamma'}^{(\phi)}$ denote the ordered products of the blocks in Eq.~\eqref{eq:coherentKraus}.  For four collisions,
\begin{equation}
 \begin{aligned}
 P_\phi(\gamma)
 &=r_n\prod_{i=1}^{4}p_{\nu_i}
 |\langle m|\mathcal K_\gamma^{(\phi)}|n\rangle|^2,\\
 P_\phi(\gamma')
 &=r_m\prod_{i=1}^{4}p_{\mu_i}
 |\langle n|\mathcal K_{\gamma'}^{(\phi)}|m\rangle|^2,\\
 \sigma_\phi(\gamma)
 &=\ln\frac{r_n}{r_m}
 +\beta\omega\sum_{i=1}^{4}(\mu_i-\nu_i).
 \end{aligned}
 \label{eq:coherentpath}
\end{equation}
The pulse contributes work but no bath entropy; the measured ancilla jumps supply the second term in $\sigma_\phi$.  The paired path laws give
$\Omega_\phi=\ln[P_\phi(\gamma)/P_\phi(\gamma')]$ and hence the dynamical term trajectory by trajectory.  We set $\theta=1.2$, $\beta\omega=1.6$, vary $\phi$ over $[0,1.3]$, and exactly enumerate the full set of
$2\times2\times4^4=1024$ endpoint-and-ancilla records throughout the sweep.  At $\phi=0$ the energy-conserving exchange channel is reversible in its thermal stationary state.  Turning on the pulse creates stationary coherence and a forward--reverse amplitude mismatch, making $\Sigma^\ast$ nonzero while leaving the theorem unchanged.  Full Kraus operators, representative values, and numerical audits of the sweep are given in the Supplemental Material~\cite{SupplementalMaterial}.

Every point in Fig.~\ref{fig:frequency}(b) is obtained by exact enumeration of the measured-record probabilities.  At $\phi=0.8$, for example, $\meanOmega=2.500$, $\Sigma=1.084$, and $\Sigma^\ast=1.416$, while $\Pi_\sigma=0.237$, $\Pi_\Omega=0.175$, and $L=0.056$.  These are tie-corrected sign statistics because the discrete record law assigns positive probability to zero-entropy events.  The physical and completed statistics are distinct, and neither is determined by $\Sigma$ alone.  More generally, the exact gap in Eq.~\eqref{eq:gap} resolves the separation into records for which $\sigma$ and $\Omega$ have opposite signs; coherent control changes this sign-misaligned sector without modifying the theorem.

\emph{Frequency--severity law.---} Although $\sigma$ need not obey the DFT in Eq.~\eqref{eq:microdft}, its definition gives the integral fluctuation theorem $\langle e^{-\sigma}\rangle=1$.  A direct application of Markov's inequality gives the standard tail bound \cite{Cavina2016,MillerPerarnauLlobet2023}, for $a>0$,
\begin{equation}
 \Prob(\sigma\le-a)\le e^{-a},
 \label{eq:tail}
\end{equation}
Equation~\eqref{eq:tail} is textbook, and Fig.~\ref{fig:mild}(a) is only a baseline.  Combining the tie-corrected floor $\Pi_\sigma\ge L(\meanOmega)$ with this standard ceiling gives the new consequence
\begin{equation}
 \Prob(-a<\sigma<0)+\frac12\Prob(\sigma=0)
 \ge[L(\meanOmega)-e^{-a}]_+.
 \label{eq:mild}
\end{equation}
When $\Prob(\sigma=0)=0$, the same combination also controls the conditional severity distribution:
\begin{equation}
 \begin{aligned}
 \Prob(-\sigma\ge a\mid\sigma<0)
 &\le\frac{e^{-a}}{L(\meanOmega)},\\
 \operatorname{median}(-\sigma\mid\sigma<0)
 &\le\ln\frac{2}{L(\meanOmega)}.
 \end{aligned}
 \label{eq:conditional}
\end{equation}
Equations~\eqref{eq:mild} and \eqref{eq:conditional} are our second main result: completed irreversibility prevents negative physical records from becoming too rare, while the physical fluctuation theorem confines most of their conditional probability to a finite neighborhood of zero.  Figure~\ref{fig:mild}(b) illustrates the first statement.  When a point mass is present at $\sigma=0$, Eq.~\eqref{eq:mild} retains its explicit half-weighted tie contribution.  The corresponding arbitrary-quantile form and derivation are given in the Supplemental Material~\cite{SupplementalMaterial}.

The result also turns sign statistics into an operational witness of dynamical asymmetry.  Let
$q_\sigma=\min\{\Pi_\sigma,1-\Pi_\sigma\}$.  Because the two-sided tie-corrected bound implies
$q_\sigma\ge L(\meanOmega)$ and $L$ decreases monotonically, the observed sign imbalance can be inverted.  Setting
$a=\ln[(1-q)/q]$ gives $q=(1+e^a)^{-1}$ and
$h(a)=a\tanh(a/2)=(1-2q)\ln[(1-q)/q]$.  Therefore, with
$\mathcal I(q)=(1-2q)\ln[(1-q)/q]$, $\mathcal I(0):=+\infty$, and $\mathcal I(1/2):=0$,
\begin{equation}
 \meanOmega\ge\mathcal I(q_\sigma),\qquad
 \Sigma^\ast\ge[\mathcal I(q_\sigma)-\Sigma]_+.
 \label{eq:witness}
\end{equation}
Thus the measured mean physical entropy and the tie-corrected sign statistic can certify a minimum hidden forward--backward dynamical-asymmetry cost, even when $\sigma^\ast$ is not reconstructed trajectory by trajectory.  Operationally, the second inequality compares a sign-only lower estimate of the completed dissipation with the independently measured mean $\Sigma$.  This is a population-level statement; finite-sample confidence bounds for empirical sign counts are left for future work.

At the level of path laws satisfying the product identity and physical IFT, $\Sigma$ alone yields no positive floor.  In the two-orbit family detailed in the Supplemental Material~\cite{SupplementalMaterial}, taking $w=\varepsilon$ and $t=1-\varepsilon$ at fixed $\Sigma>0$ gives $\Pi_\sigma=[1-(1-\varepsilon)^2]/2\to0$ and $\meanOmega=(1-\varepsilon)^2\ln[(2-\varepsilon)/\varepsilon]\to\infty$.  Thus no $\Sigma$-only floor follows from these identities without additional protocol constraints.

\paragraph{Discussion.---}
The ordering from which Eq.~\eqref{eq:main} follows separates two notions that are often conflated.  The event $\sigma<0$ is an apparent violation of the physical second law $\Sigma\ge0$.  The event $\Omega<0$ is an error by the optimal equal-prior classifier that decides whether a record was drawn from $P$ or from its reversed image $P'$.  Likelihood-ratio decision rules and their arrow-of-time interpretation are established \cite{ParrondoArrow2009,RoldanDecision2015,BenoistRepeated2018}.  Here their sign optimality yields the thermodynamic transfer: $\Pi_\Omega=(1-\|P-P'\|_{\rm TV})/2$ cannot exceed the error of the physical sign rule, so $\Pi_\sigma\ge\Pi_\Omega$ even when $\sigma$ does not satisfy a DFT.

The two upper bounds answer different questions.  The ceiling $\Pi_\Omega\le1/2$ follows because $\Omega$ obeys the DFT, exactly as in Ref.~\cite{SalazarApparent2021}.  It transfers to $\Pi_\sigma$ in protocols where $\sigma$ itself obeys a forward DFT.  At the path-law level, adjacent reversal pairs with $P=(0.72,0.08,0.18,0.02)$ and $\sigma=(-0.1,0.1,s,-s)$, $s\simeq0.657$, satisfy the physical IFT and give $\Pi_\sigma=0.74>1/2$; its explicit IFT verification is given in the Supplemental Material~\cite{SupplementalMaterial}.  Thus the $1/2$ ceiling does not follow from the stated path-law identities; a dedicated quantum-collision realization of this example remains open.  The general ceiling $\Pi_\sigma\le1-L(\meanOmega)$ remains useful: together with the lower branch it constrains either sign near reversibility and enables Eq.~\eqref{eq:witness}.

Classical fluctuation theorems determine positive-to-negative probability ratios, while martingale and stopping-time methods constrain extrema and stopped processes \cite{NeriRoldanJulicher2017,Pigolotti2017,ManzanoMartingale2019,ManzanoGambling2021}.  Here we address a fixed-horizon \emph{frequency} when the exact DFT belongs to a distinct completed variable.  Unlike generic coarse-graining anomalies \cite{EspositoCoarseGraining2012,Degunther2024,Parrondo2015}, the missing contribution is explicit, reconstructed from paired records, and quantified by $\Sigma^\ast$.

The result is not restricted to discrete collisions.  For continuous monitoring, let $\mathbb P$ be the forward path measure and $m$ an involutive record reversal.  Whenever $\mathbb P$ and $\mathbb P\circ m$ are mutually absolutely continuous,
$\Omega[\Gamma]=\ln\{\dd\mathbb P/\dd(\mathbb P\circ m)\}[\Gamma]$ obeys the same DFT by change of measure.  If a physically backward path measure provides the decomposition in Eq.~\eqref{eq:completion}, the sign theorem, frequency floor, severity bounds, and asymmetry witness carry over to jump and diffusive trajectories \cite{HasegawaContinuous2020,ManzanoZambrini2022,LandiKewming2024,HegdePottsLandi2025,Harrington2019,Rossi2020,Jayaseelan2021,Belenchia2022,FerriCortes2025}; the measure-theoretic construction is detailed in the Supplemental Material~\cite{SupplementalMaterial}.

\paragraph{Conclusion.---}
Apparent second-law violations are often presented as exceptional negative-entropy records.  Here they become a quantitatively constrained feature of quantum trajectories.  Starting from the arbitrary-coupling decomposition $\Omega=\sigma+\sigma^\ast$, we identified the completed entropy as the reversal log-likelihood and proved that its sign is optimal among all reversal-odd trajectory observables.  This transfers the sharp DFT frequency floor to the physical sign statistic even when $\sigma$ itself has no forward DFT, yielding the compact zero-inclusive inequality in Eq.~\eqref{eq:main} and its tie-corrected two-sided form.

The accompanying consequences make the theorem operational.  The physical IFT converts the frequency floor into a ``frequent but mild'' law and conditional severity bounds, while a measured sign imbalance certifies a minimum hidden mean $\Sigma^\ast$.  Random strong-coupling collisions and a coherently driven qubit show how these statements can be tested from endpoint populations, ancilla energy changes, and paired record frequencies.  The limitation is equally informative: $\Sigma$ alone cannot support a positive universal floor; the controlling cost is $\meanOmega=\Sigma+\Sigma^\ast$.  Because the argument rests on path reversal and change of measure, it also carries to continuous monitoring.  Apparent violations are therefore not merely exceptions to average irreversibility, but practical probes of the dynamical asymmetry hidden by the physical entropy alone.

\begin{acknowledgments}
OpenAI's Codex with GPT-5.6-Sol assisted with literature triage, mathematical and citation cross-checks, and LaTeX revision.  The author directed the scientific questions and revision criteria, reviewed the AI-assisted output, and takes full responsibility for the manuscript.
\end{acknowledgments}

\bibliography{references}
\end{document}